\documentclass[sort&compress,10pt,5p]{elsarticle}
\usepackage{amssymb}
\usepackage{hyperref}

\title{First Commissioning of a Cryogenic Distillation Column for Low Radioactivity Underground Argon}
\author[princeton]{H. O.~Back\corref{cor1}}
\author[fnal]{T.~Alexander}
\author[augustana]{A.~Alton}
\author[princeton]{C.~Condon}
\author[princeton]{E.~de Haas}
\author[princeton]{C.~Galbiati}
\author[princeton]{A.~Goretti}
\author[princeton]{T.~Hohmann}
\author[princeton]{An.~Ianni}
\author[fnal]{C.~Kendziora}
\author[princeton]{B.~Loer\fnref{fn1}}
\author[fnal]{D.~Montanari}
\author[princeton]{P.~Mosteiro}
\author[fnal]{S.~Pordes}

\cortext[cor1]{Corresponding author. {email: hback@fnal.gov} \\{Tel.: +1-630-840-2218}, {fax +1-630-840-4263} 
\\{address: Fermilab MS351, P.O. Box 500, Batavia, IL 60510}}
\fntext[fn1]{Now at Fermi National Accelerator Laboratory}

\address[princeton]{Department of Physics, Princeton University, Jadwin Hall, Princeton, NJ 08544}
\address[fnal]{Fermi National Accelerator Laboratory, P.O.~Box 500, Batavia, IL 60510}
\address[augustana]{Augustana College, Physics Department, 2001 South Summit Ave., Sioux Falls, SD 57197}

\renewcommand{\sup}[1]{$^{\mathrm{#1}}$}
\newcommand{\sub}[1]{$_{\mathrm{#1}}$}
\newcommand{\iso}[2]{\sup{#1}$\!${#2}}
\newcommand{\E}[1]{$\times10^{#1}$}
\newcommand{\ar}{\iso{39}{Ar}}

\newcommand{\imgwidth}{\columnwidth}

\begin{document}

\begin{abstract}
We report on the performance and commissioning of a cryogenic distillation column for low radioactivity underground argon at Fermi National Accelerator Laboratory. The distillation column is designed to accept a mixture of argon, helium, and nitrogen and return pure argon with a nitrogen contamination less than 10 ppm. In the first commissioning, we were able to run the distillation column in a continuous mode and produce argon that is 99.9\% pure. After running in a batch mode, the argon purity was increased to 99.95\%, with 500 ppm of nitrogen remaining. The efficiency of collecting the argon from the gas mixture was between 70\% and 81\%, at an argon production rate of 0.84-–0.98 kg/day. 
\end{abstract}


\maketitle

\section{Introduction}
Argon is a powerful scintillator and an excellent medium for detection of ionization. Argon derived from the atmosphere contains a small fraction of the radioactive isotope \ar, which undergoes beta decay (Q=565 keV, t\sub{1/2}=269 y), resulting in  a specific activity of $\sim$1~Bq/kg of atmospheric argon~\cite{Loosli1983, Benetti2007}. Background from \ar\ decays can limit the sensitivity of liquid argon scintillation detectors searching for direct dark matter interactions, and the pile-up of \ar\ events sets a limit on the order of a few hundred kilograms on the maximum practical size of two-phase argon time projection chamber (TPC) dark matter searches.

Since 2009, a production plant (the VPSA plant) has been extracting low radioactivity argon from CO\sub2 wells in a Kinder Morgan facility in southwestern Colorado~\cite{Back2012}. The underground argon from this plant has an \ar\ concentration less than 0.65\% of the \ar\ concentration in atmospheric argon~\cite{Xu2012}. The output of this plant is a crude mixture of argon, helium, and nitrogen, with an argon concentration of 3--5\%. The crude argon therefore requires further purification to produce detector-grade argon with ppb levels of impurities.

A common and effective method to separate gases is through cryogenic fractional distillation. We describe the commissioning of a cryogenic distillation column that has been constructed at Fermilab for the purification of the gas extracted from the CO\sub2 wells in Cortez, CO.

\section{Cryogenic fractional distillation}
It is well understood that the difference in volatility of the constituents in a multi-component fluid allows for
separation through distillation~\cite{Arey1900}.  It is possible to perform distillation continuously in a column packed with a high-surface-area material and with a temperature gradient maintained along its length~\cite{McCabe1976,Perry1997,Halvorsen1999}. The liquid will boil and the gases will condense continuously on the packing material in the column; gases rise and recondense, while liquids sink and reboil. The components with higher volatility rise preferentially and components with lower volatility fall to the lower volume (the reboiler). By maintaining a temperature gradient and a constant flow of liquid into the column, an equilibrium is established, and very pure material can be collected continuously from the column.

\begin{figure}
  \centering
 \includegraphics[width=0.7\columnwidth]{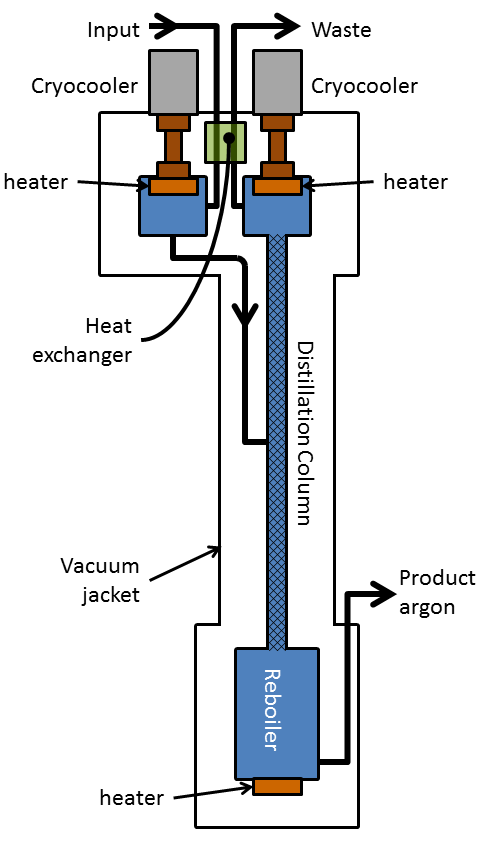}
 \caption{Schematic of the cryogenic distillation column at Fermilab.}
 \label{fig:columnschematic}
\end{figure}

The design of the distillation system is based on the McCabe-Thiele (M-T) method for the design and analysis of distillation column systems~\cite{McCabe1976,Perry1997,Halvorsen1999}. The column is a 318 cm long stainless steel tube filled with a packing material (Sulzer Chemtech, EX Laboratory Packing) making it equivalent to 58 distillation stages. The argon and nitrogen portions of the crude gas produced by the VPSA plant in Colorado are liquefied as they enter the distillation system and injected into the center of the column. The temperature gradient required along the column is maintained by a cryocooler at the top of the column balanced by heaters at the top of the column and on the reboiler at the bottom of the column. The more volatile waste gases (N\sub2 and He) are expelled from the top of the column, while the pure argon collects in the reboiler. Figure~\ref{fig:columnschematic} shows a schematic overview of the distillation column.

In order to monitor the distillation process, we measure the composition of the gas feed (input), the gas waste, and the liquid that is collected in the reboiler (product), with a Stanford Research System (SRS) Universal Gas Analyzer (UGA). UGAs require calibration for each species of gas they are measuring; they are not able to make quantitative determinations of the constituent gas fractions of an unknown gas.
The UGA essentially measures the current of ionized gas at a given mass-to-charge ratio, and then reports this as a partial pressure for a given mass~\cite{SRSA}. If a molecule or atom is doubly ionized, it will contribute to the signal at half the true mass. Molecules and atoms having the same measured masses will interfere in the measurements. The UGA is calibrated to nitrogen, and therefore the partial pressure reported by the UGA for nitrogen (mass 28) is correct~\cite{SRSA}. Our gas is a mixture of argon (mass 40), helium (mass 4), and nitrogen (mass 28), and these three gases do not have any overlapping mass-to-charge ratios, hence no interferences. Therefore we use a known mixture of argon, helium, and nitrogen to calibrate the argon and helium using the partial pressure of nitrogen.

Our gas mixture also allows for useful relative gas measurements, independent of calibration. Helium cannot be collected in the distillation column because of its low boiling point, and essentially all helium that enters the column leaves through the waste outlet. This allows us to determine how much of the input argon and nitrogen are being captured by the column, by measuring the ratio of argon to helium and nitrogen to helium at the input, and comparing them with the same ratios in the waste.

\begin{figure}
 \centering
 \includegraphics[width=\imgwidth]{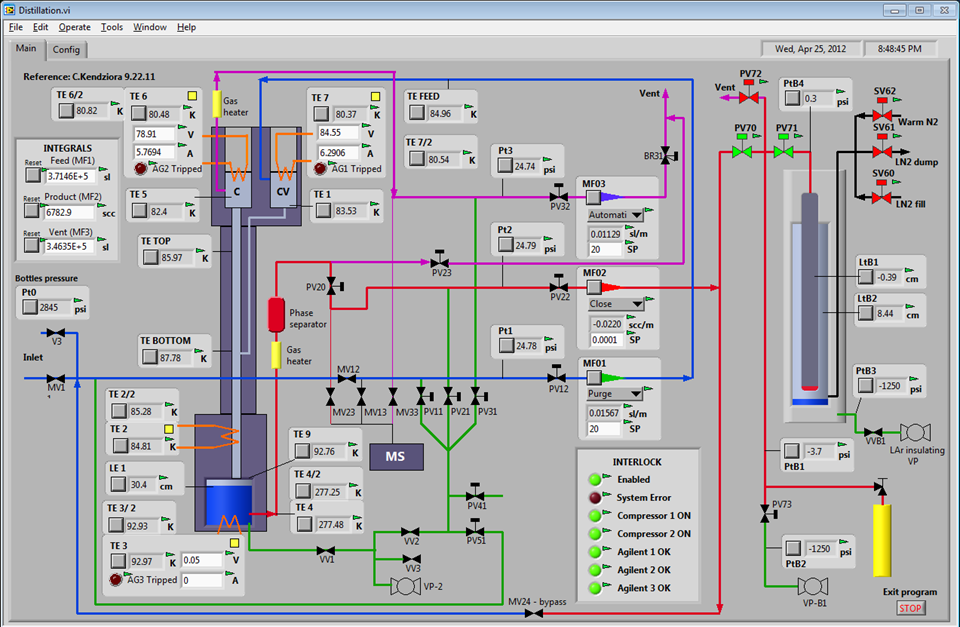}
 \caption{LabVIEW control system GUI.}
 \label{fig:labviewscreen}
\end{figure}

As illustrated in Figure~\ref{fig:columnschematic}, cooling for the column operation is provided by two cryocoolers (Cryomech Al600 Cryorefrigerators), which run constantly at full power, and the temperatures along the column are set by heaters (controlled by LakeShore Model 336 Cryogenic Temperature Controllers). The column control and data acquisition program is written in LabVIEW running on a National Instruments CompactRIO platform. Figure~\ref{fig:labviewscreen} shows the control screen.

\section{Commissioning}
To commission the cryogenic distillation column, we obtained a known mixture of gas that is approximately the same as the output of the VPSA plant in Colorado. Using this known gas, the UGA was calibrated to measure argon, helium, and nitrogen. Table~\ref{tab:ugacalibration} shows the gas mixture used and the calibration factors obtained for the UGA.

\begin{table}
  \caption{Composition of the gas used to calibrate the UGA, and the resulting calibration factors. The UGA must be calibrated to the nitrogen partial pressure in a known mixture of gas. }
  \label{tab:ugacalibration}
  \centering
  \begin{tabular}{llllll}
    \hline
    \hline
    Gas & Conc. & \multicolumn{2}{c}{Pressure (torr)} & Calib. \\
    (AMU) & & Measured & Calculated & Factor \\
    \hline
    N\sub2 (28) & 40\% & 2.15\E{-6} & 2.15\E{-6} & 1.00 \\
    Ar (40)     & 5\%  & 3.90\E{-7} & 2.69\E{-7} & 0.69 \\
    He (4)      & 55\% & 2.16\E{-6} & 2.96\E{-6} & 1.37 \\
    \hline
  \end{tabular}

\end{table}

\begin{figure}
  \centering
  \includegraphics[width=\imgwidth]{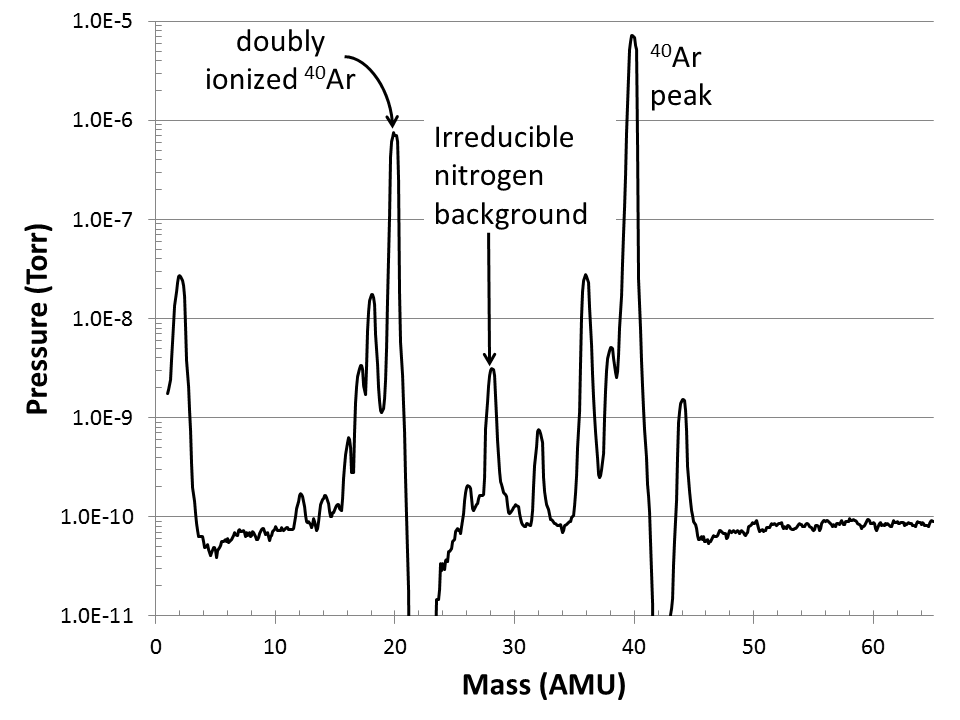}
  \caption{Pressure vs.\ mass spectrum of a pure argon sample, showing the irreducible nitrogen background peak at mass 28. This background puts a lower limit on the sensitivty of the UGA to measure nitrogen at $\sim$500~ppm.}
  \label{fig:ugabackground}
\end{figure}

A sample of pure argon, with nitrogen content below 1~ppm, was used to determine the sensitivity limit of the UGA to measure nitrogen in pure argon. In the pressure versus mass spectrum in Figure~\ref{fig:ugabackground}, there is a clear peak at mass 28. When the calibration factor is taken into account for the argon peak, the ultimate sensitivity to nitrogen in argon is found to be $\sim$500 ppm.

The cryogenic distillation column can operate in two modes. It is designed to operate in a continuous distillation mode, where the gas to be separated is fed into the column continuously, while pure argon is collected in the reboiler, and nitrogen and helium are exhausted through the waste. However, if the conditions required for continuous flow operations do not result in adequate purity of the argon collected in the reboiler, the distillation column can be operated in a batch purification mode. In this mode, the input is turned off, and the liquid in the reboiler is further distilled with a retuned column temperature profile.

The distillation column was initially operated in the continuous flow mode. The temperatures of the distillation column were tuned to maximize the amount of argon collected, by minimizing the argon in the waste.
\[ \left(\frac{\mathrm{Ar}}{\mathrm{He}}\right)_{input} \gg \left(\frac{\mathrm{Ar}}{\mathrm{He}}\right)_{waste}\]
At the same time, we wish to minimize the amount of nitrogen contamination in the reboiler, effectively by maximizing the amount of nitrogen in the waste.
\[ \left(\frac{\mathrm{N_2}}{\mathrm{He}}\right)_{input} \approx \left(\frac{\mathrm{N_2}}{\mathrm{He}}\right)_{waste}  \]
As mentioned, normalizing to the helium measurement avoids the need to rely on a specific argon calibration of the UGA. 

\begin{figure}
  \centering
  \includegraphics[width=\imgwidth]{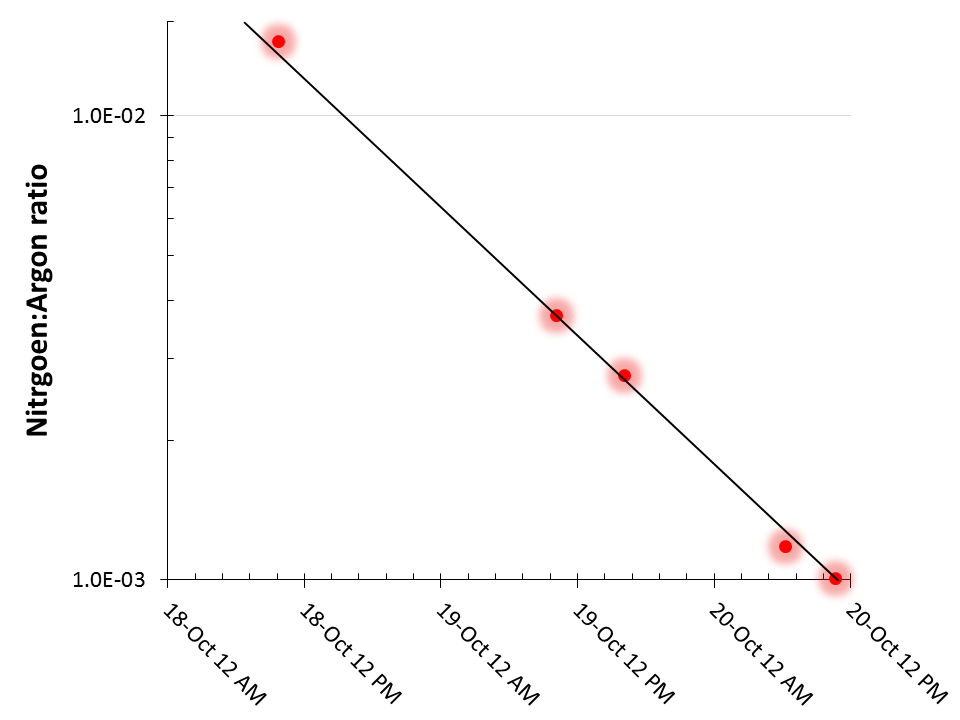}
  \caption{Nitrogen:argon ratio as a function of time measured in the reboiler during continuous distillation.}
  \label{fig:argonfraction}
\end{figure}

After a volume of liquid has been collected in the reboiler, we can measure the gas in the argon product line to determine the amount of nitrogen in the collected liquid. Figure~\ref{fig:argonfraction} shows the nitrogen:argon ratio of the product gas coming from the reboiler. As is clear, the nitrogen concentration decreased continuously until the trial was stopped --- because the input feed gas supply was consumed.

\begin{figure}
  \centering
  \includegraphics[width=\imgwidth]{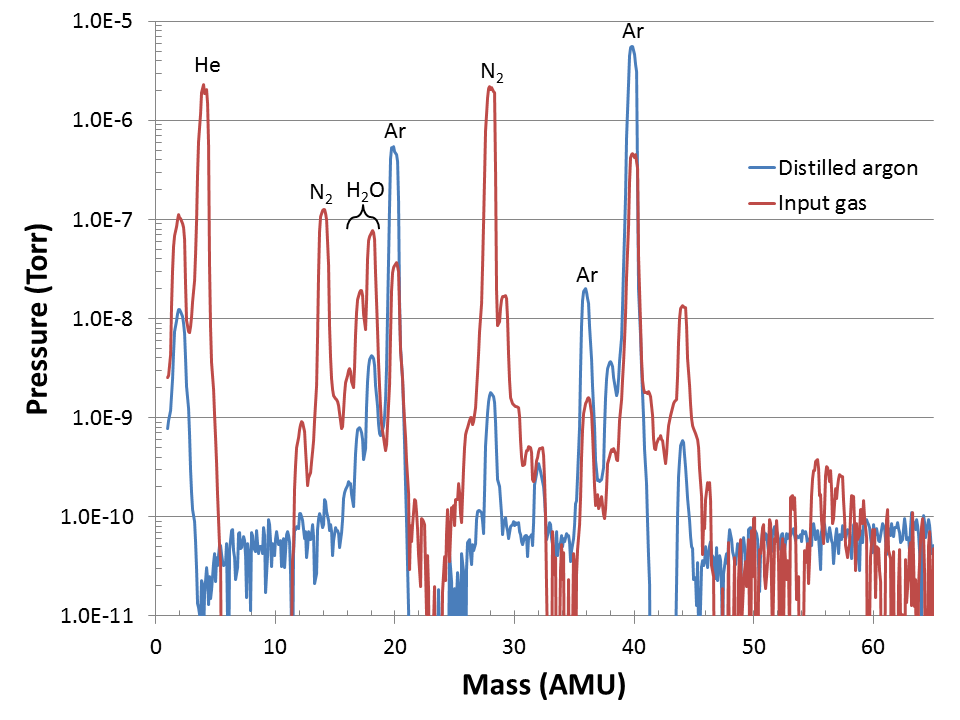}
  
  \caption[blank]{UGA spectra of the input gas mixture and the distilled argon product during commissioning.\footnotemark[2]} 
  \label{fig:continuous_partials}
\end{figure}

The final nitrogen concentration achieved in the continuous flow mode before the gas was consumed was $\sim$1000 parts per million (ppm), giving 99.9\% pure argon. This measurement was confirmed by 2 independent measurements of a sample of the gas: Atlantic Analytical Laboratory reported that the sample contained 700 ppm of nitrogen, and colleagues at Pacific Northwest National Laboratory measured the nitrogen content to be 920 ppm. The data from the UGA show that the nitrogen concentration was decreasing throughout the continuous distillation phase, and we are confident that continuous distillation can produce argon with a nitrogen contamination well below 1000 ppm. Figure~\ref{fig:continuous_partials} shows a comparison of the pressure versus mass spectra of the input gas and of the pure argon produced by the cryogenic distillation column. As can be seen in the figure, the nitrogen and helium peaks are both reduced by many orders of magnitude, leaving essentially pure argon. 

Since the feed gas can vary in its argon concentration, which will affect the performance and optimal tuning of the column, we tested the distillation column on an input with concentrations different from those given in Table~\ref{tab:ugacalibration}, by adding a final cylinder of gas with a mixture of 2.5\% Ar, 27.5\% He, and 70\% N\sub2. Almost immediately, the purity of the argon in the reboiler was compromised with a higher concentration of nitrogen. This eventually resulted in a N\sub2 concentration of 10\%, implying that, rather than sending essentially all of the input nitrogen to the waste output, we were only sending 80\% to the waste. This showed that the operating parameters for the distillation column must be tuned for different input gas compositions in continuous mode.

\begin{figure}
  \centering
  \includegraphics[width=\imgwidth]{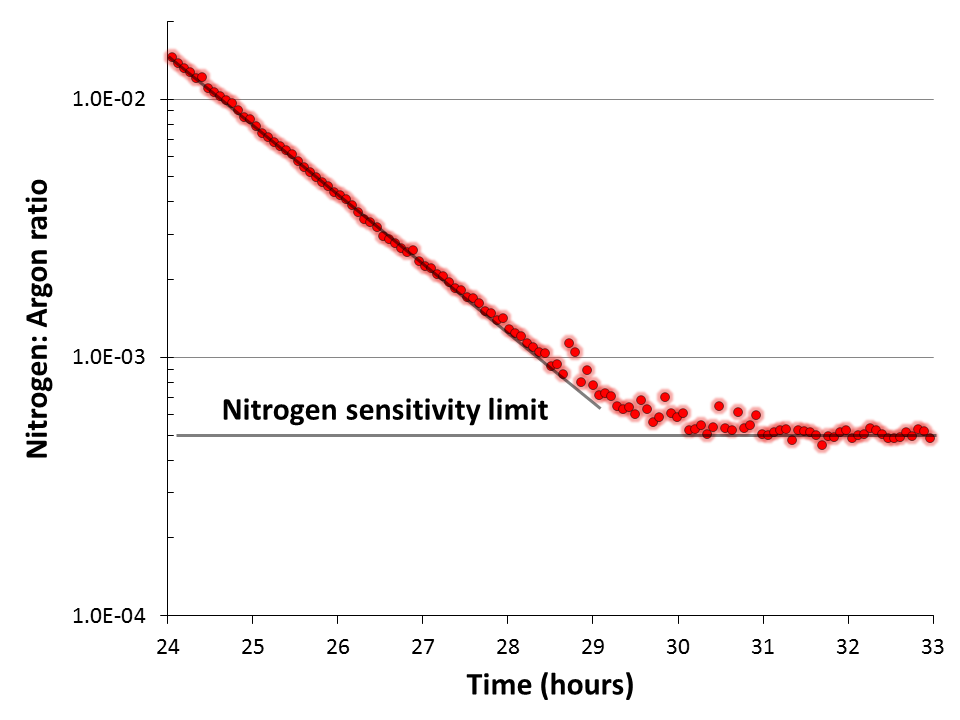}
  \caption{Nitrogen/argon ratio as a function of time during batch purification.}
  \label{fig:nfrac_batch}
\end{figure}

With this high concentration of nitrogen in the argon in the reboiler, the batch distillation technique was tested. In this mode, the input stream is turned off, and the temperature gradient along the column was retuned to allow the excess  nitrogen to escape the reboiler, while preserving the argon.
Over several hours, the measured nitrogen concentration decreased until the nitrogen sensitivity limit of the UGA was reached. As mentioned, the lower limit of the UGA's sensitivity to measure the nitrogen concentration is $\sim$500~ppm, which is equivalent to 99.95\% pure argon. Figure~\ref{fig:nfrac_batch} clearly shows the measured nitrogen concentration decreasing and plateauing at the nitrogen sensitivity limit of the UGA.

\footnotetext[2]{The differing water peaks are due to sampling conditions, and are not part of the sampled gases.}

\section{Collection efficiency}
In addition to achieving high purity, it is important that a minimum of the argon in the feed gas be wasted. We can calculate the amount of argon that was fed into the distillation from the total amount of gas consumed. We consumed 24 high-pressure cylinders of the gas mixture stated in Table~\ref{tab:ugacalibration}.  Each high-pressure cylinder contains 7419~std.\ liters of gas, which corresponds to a total of 14.9~kg of argon in the 24 cylinders combined.

It is possible to estimate the amount of argon collected by two independent methods. The most straightforward is from the liquid level recorded in the reboiler. This provides the liquid volume of the argon, and therefore the total mass. The other method is to measure with the mass flow meter the integral amount of boil off argon gas that is released from the reboiler when it is warmed to room temperature.
The flow of argon required to allow the UGA to sample the gas from the reboiler throughout the commissioning run must also be accounted for, and was about 1.2~kg. 

Table~\ref{tab:collectionefficiency} shows the masses calculated by both methods for determining the total argon collected.  The integrated efficiency for capturing the argon from the input gas mixture was between 70\% and 81\%. The discrepancy may be attributed to inaccurate calibrations of either the liquid level monitor or the mass flow controller.

\begin{table}
  \caption{Calculated argon collection efficiency, using 2 independent methods. The total input mass of argon was $\sim$14.9~kg, with $\sim$1.2~kg wasted due to UGA sampling.}
  \label{tab:collectionefficiency}
  \centering
  \begin{tabular}{l l l}
  \hline \hline
  {Method} & {Calculated Mass} & {Collection Efficiency} \\
  \hline
  Liquid level & 12 kg & 81\% \\
  Mass flow & 10.5 kg & 70\% \\
  \hline
  \end{tabular}
\end{table}

\section{Argon production rate}
Another important parameter for the distillation column is the overall production rate of purified argon. The input gas was fed into the column at a constant rate of 10~sL/m. When we take into account that the mixture only contains 5\% argon and the collection efficiency is between 70\% and 81\%, (Table~\ref{tab:collectionefficiency}), we find that the production rate of argon is in the range of 0.84--0.98 kg/day.

\section{Conclusion}
In the first commissioning of the cryogenic distillation column at Fermilab, we have shown that it can effectively reduce the nitrogen content by more than 3 orders of magnitude and helium by more than 5 orders of magnitude. The argon produced by the distillation column contains less than 500~ppm of nitrogen, and the helium has effectively been eliminated. This argon purification was performed at a rate of 0.84--0.98~kg/day with 75$\pm$5\% collection efficiency.
With this commissioning phase complete, we have now started to operate the distillation system to produce high-purity, low-radioactivity underground argon.


\section{Acknowledgements}
This work was supported in part by National Science Foundation grants PHY NSF-0704220, PHY NSF-0811186, and PHY NSF-1004072 and by Fermilab under DOE Contract \# De-AC02-07CH11359.

Support for Henning Back at Princeton University was provided by Mark Boulay and Art McDonald of Queen's University, Kingston, Canada, through grants from the Canada Foundation for Innovation and the Ministry of Research and Innovation of the Province of Ontario.

We would like to thank Stanley Bos, Craig Aalseth, and John Orrell for performing the analysis of the gas sample of the continuous distillation phase.

Special thanks are due to the the technical staff at the Proton Assembly Building at Fermilab: Ron Davis, Kelly Hardin, Walter Jaskierny, William Miner, and Mark Ruschman.

We would also like to acknowledge the efforts of undergraduate students: Josh Bonnat from University of Massachusetts, Amherst and Hannah Rogers from Augustana College Sioux Falls.

\end{document}